# Coalition Structure Generation over Graphs

**Thomas Voice**                                    TDV@ECS.SOTON.AC.UK
*School of Electronics and Computer Science,*
*University of Southampton, UK*

**Maria Polukarov**                                 MP3@ECS.SOTON.AC.UK
*School of Electronics and Computer Science,*
*University of Southampton, UK*

**Nicholas R. Jennings**                            NRJ@ECS.SOTON.AC.UK
*School of Electronics and Computer Science,*
*University of Southampton, UK*
*Department of Computing and Information Technology,*
*King Abdulaziz University, Saudi Arabia*

## Abstract

We give the analysis of the computational complexity of *coalition structure generation over graphs*. Given an undirected graph $G = (N, E)$ and a valuation function $v : \mathcal{P}(N) \to \mathbb{R}$ over the subsets of nodes, the problem is to find a partition of $N$ into connected subsets, that maximises the sum of the components' values. This problem is generally NP–complete; in particular, it is hard for a defined class of valuation functions which are *independent of disconnected members*—that is, two nodes have no effect on each other's marginal contribution to their vertex separator. Nonetheless, for all such functions we provide bounds on the complexity of coalition structure generation over general and minor–free graphs. Our proof is constructive and yields algorithms for solving corresponding instances of the problem. Furthermore, we derive linear time bounds for graphs of bounded treewidth. However, as we show, the problem remains NP–complete for planar graphs, and hence, for any $K_k$ minor–free graphs where $k \geq 5$. Moreover, a 3-SAT problem with $m$ clauses can be represented by a coalition structure generation problem over a planar graph with $O(m^2)$ nodes. Importantly, our hardness result holds for a particular subclass of valuation functions, termed *edge sum*, where the value of each subset of nodes is simply determined by the sum of given weights of the edges in the induced subgraph.

## 1. Introduction

*Coalition structure generation* (CSG) is the equivalent of the complete set partitioning problem (Yeh, 1986)—one of the fundamental problems in combinatorial optimisation, that has applications in many fields, from political sciences and economics, to operations research and computer science. In a CSG problem, we have a set $N$ of $n$ elements and a valuation function $v : \mathcal{P}(N) \to \mathbb{R}$, where $\mathcal{P}(N)$ denotes the power set of $N$, and the problem is to divide the given set into disjoint exhaustive subsets (or, *coalitions*) $N_1, \ldots, N_m$ so that the total sum of values, $\sum_{i=1}^{m} v(N_i)$, is maximised. Thus, we seek a most valuable partition (or, a *coalition structure*) over $N$.

Partitioning structure problems arise in a wide range of practical domains including delivery management, scheduling, routing and location problems, where one wishes to assure that every customer is served by one (and only one) location, vehicle or person (server). Commonly cited problems of this kind include the *crew-scheduling problem* where every flight leg of an airline must be scheduled by exactly one cockpit crew, the *political districting problem* whereby regions must be divided into voting districts such that every citizen is assigned to exactly one district, and the *coalition formation problem* of political parties (Balas & Padberg, 1976). Recently, CSG has become a major research topic in artificial intelligence and multi-agent systems, as a tool for autonomous agents to form effective teams. For example, in electronic commerce buyer agents may pool their demands in order to obtain group discounts (Tsvetovat, Sycara, Chen, & Ying, 2001); in e-business coalitions may form in order to satisfy certain market niches as they can respond to more diverse orders than individual agents (Norman, Preece, Chalmers, Jennings, Luck, Dang, Nguyen, Deora, Gray, & Fiddian, 2004); and in distributed vehicle routing coalitions of delivery companies can reduce the transportation costs by sharing deliveries (Sandholm & Lesser, 1997). Other important applications include information gathering where several information servers come together to answer queries (Klusch & Shehory, 1996), multi-sensor networks where sensors form dynamic coalitions in wide-area surveillance scenarios (Dang, Dash, Rogers, & Jennings, 2006), and grid computing where multi-institution virtual organisations are viewed as being central to coordinated resource sharing and problem solving (Yong, Li, Weiming, Jichang, & Changying, 2003).

However, the classic CSG model assumes no structure on the primitive set of elements. This is a considerable shortcoming, as in various contexts of interest to computer scientists, these elements represent agents (either human or automated) or resources (e.g., machines, computers, service providers or communication lines), which are typically embedded in a social or computer network. Moreover, in many such scenarios those elements which are disconnected have no effect on each other's performance and potential contribution to a coalition, and if not connected by intermediaries, may not be able to cooperate at all. For example, consider a *communication network* where each edge is a channel, with capacity indicating the amount of information that can be transmitted through it. Thus, in the aforementioned contexts of e-commerce, multi-sensor networks or grid computing, such a network connects between sellers and buyers, sensors or agents working on computational tasks, respectively. Any subset of nodes in this network produces a value proportional to the total capacity of the subnetwork induced by these nodes. In such a scenario, any two nodes that are not connected by a direct link in the network, will not affect each other's marginal contribution to any coalition of nodes that separates them. Or, as is also typical in e-commerce and e-business domains, assume that an edge represents a trust link in a *reputation system*, so that two nodes will only participate in the same coalition if the trust distance given by the length of a path between them, is finite (that is, a coalition induces a connected subgraph of the trust network). Suppose that a value of a coalition is given by the number of pairs of its mutually trusted members—i.e., the edges in the induced subgraph. Then, a contribution of a particular node $i$ will not depend on another node $j$ who trusts some members of the coalition but does not trust $i$ directly, as there is no edge between $i$ and $j$. Additional natural examples arise in *multi-agent systems* domains, where agents come together to complete tasks. Typically, a pair of agents can be associ-ated with a "weight" indicating their potential mutual (in)efficiency in the task execution (e.g., due to skill/expertise or equipment complementarity, interpersonal (in)compatibility, (dis)agreements, spatial or other constraints). The value of a coalition is then measured by the total coalitional weight as given by the sum of weights on the links whose both ends participate in the coalition. Importantly, these weights can be positive or negative, representing different relations among the agents, and thus having corresponding effects on a coalitional value. Note that agents with zero weight links do not affect each other's contribution to a coalition. Finally, *correlation clustering*—a well-known clustering technique motivated by the problem of clustering a large corpus of objects, such as documents (e.g., web pages and weblog data with given content/access patterns), customers and service providers (with given properties and past buying/selling records) or biological species (plants and animals given their features)—operates in a setting where the elements which need to be partitioned into clusters (by topic, location, behaviour etc.) are characterised by "similarity" (and/or "difference") relations among them. The aim is usually to maximise the overall agreement—i.e., correlation—of clusters. For example, given a signed graph where the edge label indicates whether two nodes are similar (+) or different (−), the task is to cluster the nodes so that similar objects are grouped together, and different ones—separately. Thus, the value of a cluster $C$ is given by the total sum of its positive intra-cluster edges and negative inter-cluster edges with one end in $C$. In such cases, only connected (either positively or negatively) members have an impact on the cluster values.

Against this background, in this paper we extend the CSG problem to connected sets. More precisely, we introduce the independence of disconnected members and consider coalition structures over the node set of a graph, endowed with a valuation function that has this property. This is formally defined in Section 2 below, where we also give necessary graph–theoretic notation and summarise our main contributions. Then, in Sections 3, 4 and 5, we discuss our results in great detail and present all the proofs. Specifically, Section 3 provides computational bounds on coalition structure generation over general graphs, and Section 4 introduces our technique for solving the problem using tree decompositions. This technique, in particular, allows us to show linear time solvability for graphs with bounded treewidth. In Section 5, we apply it to derive upper bounds for graphs with separator theorems and, in particular, planar graphs and minor–free graphs. We also present our negative result showing the NP–hardness of the problem over planar graphs and hence, all $K_k$ minor–free graphs, even for a simple, so called "edge sum", valuation function. We discuss the related literature in Secion 6. Finally, Section 7 concludes the paper.

## 2. Coalition Structure Generation over Graphs

In this section, we formalise the concepts of independence of disconnected members and graph coalition structure generation, and list our main contributions. For completeness, we first provide some graph–theoretic definitions and notation necessary for presentation of our results in following sections.

### 2.1 Graph–Theoretic Definitions and Notation

Let $N$ be a set of elements and let $\mathcal{P}_k(N)$ stand for the set of all $k$-element subsets of the set $N$. A *simple undirected graph* $G$ is a pair $G = (N, E)$ where $N$ is a finite set of elements, called the *vertices* (or, *nodes*) of $G$, and $E$ is a subset of $\mathcal{P}_2(N)$—i.e., $E$ is a collection of two-element subsets of $N$ representing connections between nodes, called the *edges* of $G$.

A *complete* graph is a graph in which each pair of nodes is connected by an edge. The complete graph with $n$ nodes is denoted $K_n$. A graph $G$ is a *bipartite* graph if its vertices can be divided into two disjoint sets $N_1$ and $N_2$ such that every edge connects a vertex in $N_1$ to one in $N_2$. A complete bipartite graph, $G = (N_1 \cup N_2, E)$, is a bipartite graph such that for any two vertices, $n_1 \in N_1$ and $n_2 \in N_2$, $\{n_1, n_2\}$ is an edge in $G$. The complete bipartite graph with $|N_1| = m$ and $|N_2| = n$, is denoted $K_{m,n}$.

An undirected graph $H$ is called a *minor* of the graph $G$ if $H$ can be obtained from $G$ by a series of vertex deletions, edge deletions and/or edge contractions (removing an edge from a graph while simultaneously merging together the two vertices it used to connect). A graph $G$ is $H$ *minor–free* if $H$ is not a minor of $G$. A graph $G$ is *planar* if it is $K_5$ minor–free and $K_{3,3}$ minor–free. An important property of a planar graph is that it can be embedded in the plane, i.e., it can be drawn in such a way that no edges cross each other. A familiar special case of planar graphs is the class of *grids*: in a finite grid graph, the vertices are associated with two indices $1 \le i \le r$ and $1 \le j \le c$, and there is an edge connecting each node $n_{i,j}$ to nodes $n_{i+1,j}$ and $n_{i,j+1}$ (if such exist)—thus, there are $r$ "rows" and $c$ "columns" in such a graph, and the number of nodes is $n = rc$.

A subgraph $H$ of the graph $G$ is *induced* if for any pair of nodes $x$ and $y$ of $H$, $\{x, y\}$ is an edge of $H$ if and only if it is an edge of $G$. In other words, $H$ is an induced subgraph of $G$ if it has exactly the edges that appear in G over the same vertex set. If the vertex set of $H$ is the subset $S \subseteq N$ of the vertex set of $G$, then $H$ can be said to be induced by $S$.

A *path* in a graph is a sequence of nodes such that from each node there is an edge to the next node in the sequence, and a path is called *simple* if it contains no repeated nodes. A graph is said to be *connected* if there is a path between every pair of nodes in the graph. A *tree* is a graph in which any two nodes are connected by exactly one simple path.

Many algorithms on graphs become easy if the input graph is a tree or "tree-like". The notion of being tree-like can be formalised using the concept of treewidth: if the treewidth of a graph is small, then it is tree-like—in particular, a tree has treewidth 1. Treewidth is defined using the concept of tree decomposition—a mapping of a graph into a tree. Formally, a *tree decomposition* of $G = (N, E)$ is a pair $(X, T)$, where $X = \{X_1, \ldots, X_m\}$ for $m \le n = |N|$ is a family of subsets of $N$, and $T$ is a tree whose nodes are the subsets $X_i$, satisfying the following properties: (i) the union of all sets $X_i$ equals $N$—that is, each graph vertex is associated with at least one tree node; (ii) for every edge $\{x, y\}$ in the graph, there is a subset $X_i$ that contains both $x$ and $y$; (iii) if $X_i$ and $X_j$ both contain a vertex $x$, then all nodes $X_k$ of the tree in the (unique) path between $X_i$ and $X_j$ contain $x$ as well—i.e., the nodes associated with vertex $x$ form a connected subset of $T$ (equivalently, if $X_i$, $X_j$ and $X_k$ are nodes, and $X_k$ is on the path from $X_i$ to $X_j$, then $X_i \cap X_j \subseteq X_k$). The *width* of a tree decomposition is the size of its largest set $X_i$ minus one. Finally, the *treewidth* of a graph $G$ is the minimum width among all possible tree decompositions of $G$.

Given this notation, we can now formally define the problem of coalition structure generation over graphs.

## 2.2 Model

Recall that a coalition structure over a set of elements $N$ is defined by a collection of its disjoint exhaustive subsets $N_1, \ldots, N_m$ where $N_i \cap N_j = \emptyset$ for all $1 \leq i, j \leq m$ and $\cup_{i=1}^m N_i = N$. Given the setting with a finite set of elements $N$ in a connected undirected graph $G = (N, E)$ and a coalition valuation function $v : \mathcal{P}(N) \to \mathbb{R}$ over subsets of $N$, where $v(\emptyset) = 0$, we consider a class of coalition structure generation problems over $N$. Accordingly, we make the following definitions.

**Definition 1** *For a graph $G = (N, E)$, a function $v : \mathcal{P}(N) \to \mathbb{R}$ is* independent of disconnected members *(IDM) if for all $i, j \in N$ with $(i, j) \notin E$, and coalition $C$ with $i, j \notin C$,*

$$v(C \cup \{i\}) - v(C) = v(C \cup \{i, j\}) - v(C \cup \{j\}).$$

This means that agent $i$ contributes to a coalition $C$ exactly the same amount as to a coalition $C \cup \{j\}$ if $i$ and $j$ are not directly connected. That is, the presence of agent $j$ does not affect the marginal contribution of agent $i$ to a separating coalition. Note that Definition 1 generally does not restrict the effects the agents may have on each other if they are connected.

To give an example, suppose that each edge $\{i, j\} \in E$ is associated with a constant weight $v_{i,j} \in \mathbb{R}$. Then, the coalition valuation function

$$v(C) = \sum_{\{i,j\} \in E : i,j \in C} v_{i,j}$$

has the IDM property. We shall term such a function an *edge sum* coalition valuation function. This function is important as it naturally arises in many application scenarios (e.g., communication networks, information and multi-agent systems) and has simple representation. In the work of Deng and Papadimitriou (1994), this function is studied in the context of complexity of cooperative game-theoretic solution concepts.

Other functions of this type arise in some familiar clustering settings. For example, suppose that each edge $\{i, j\}$ is labeled by $+$ or $-$ depending of whether $i$ and $j$ have been deemed to be similar or different. For a coalition (or, cluster) $C \subseteq N$, let $E^+(C) = \{\{i, j\} = + \,|\, i, j \in C\}$ denote the set of its positive intra-cluster edges, and let $\bar{E}^-(C) = \{\{i, j\} = - \,|\, i \in C, j \notin C\}$ be the set of negative inter-cluster edges with one end in $C$. Then, the *correlation* coalition valuation function defined as

$$v(C) = |E^+(C)| + |\bar{E}^-(C)|$$

satisfies the IDM condition. Note that this function takes into account both *intra-* and *inter*coalitional connections, and thus is different from the edge sum, which only considers intracoalitional links. Maximising the sum of coalitional values over all coalition structures, produces a partition of the nodes that agrees as much as possible with the edge labels. This objective is pursued in the paper by Bansal, Blum and Chawla (2003) where they show NP-completeness of the problem over complete graphs and provide several approximation results.

Yet another example of an IDM function is found in multi-agent scenarios where coalitions of agents work on different parts of a global project. In such settings, members of a coalition must make joint decisions and communicate them to other coalitions of agents to coordinate their actions. Furthermore, when collaboration and communication is possible only between closely connected agents, it is important that the coalition includes agents who have mutual neighbours outside the coalition, so that decisions can be made and coordinated with other coalitions. Given this, the coalition valuation function

$$v(C) = \sum_{i \in C} n_i(C)$$

where $n_i(C)$ is the number of agent pairs $(j, k) \in N \times N$ so that $j \in C$, $k \notin C$ and $\{i, j\}, \{i, k\} \in E$, has the IDM property. We shall term this function a *coordination* coalition valuation function. Obviously, by considering intercoalitional links, this function is different from the edge sum. However, note also the difference between the coordination and the correlation functions. By the latter, the effect of a link between any two agents on the value of a coalition is determined by the link label and by whether or not both of these agents belong to the coalition. In contrast, the coordination function accounts in fact for 3-agent cliques, where two agents are members of the coalition and one is an outsider.

Our analysis, however, is not restricted to a particular valuation function but rather covers the class of functions characterised by Definition 1. We define a *graph coalition structure generation* (GCSG) problem as follows.

**Definition 2** *Given a connected undirected graph $G = (N, E)$ and a coalition valuation function $v : \mathcal{P}(N) \to \mathbb{R}$ which is independent of disconnected members, the* graph coalition structure generation *problem over $G$ is to maximise $v(\mathcal{C}) = \sum_{C \in \mathcal{C}} v(C)$ for $\mathcal{C}$ a coalition structure over $N$.*

GCSG can be posed as a clustering or a graph partitioning problem where the sum of cluster values, which are given by some IDM valuation function, is to be maximised. For instance, the aforementioned correlation clustering is a special case of GCSG. Note, however, that clustering problems in general do not necessarily fit in our model: indeed, some of them have objectives that do not admit the IDM property; on the other hand, some clustering problems have additional restrictions on feasible graph partitions. For example, one of the natural objectives in this domain is to maximise the *modularity* of clusters (Brandes, Delling, Gaertler, Görke, Hoefer, Nikoloski, & Wagner, 2008) given by the sum of cluster values defined as follows. For each cluster $C$, let $v(C) = \frac{|E(C)|}{|E|} - \left(\frac{|E(C)| + \bar{E}(C)|}{2|E|}\right)^2$, where $E(C) = \{\{i, j\} \in E : i, j \in C\}$ is the set of intra-cluster edges of $C$ and $\bar{E}(C) = \{\{i, j\} \in E : i \in C, j \notin C\}$ is the set of its inter-cluster edges. Notice that the second term of the valuation funciton is squared, which implies the violation of the IMD property. Another related setting is the *weighted graph partitioning* problem where nodes and edges have (non-negative) weights and the aim is to divide the graph into $k$ disjoint parts such that the parts have approximately equal weight and the size of the edge cut is minimised. Crucially, unlike in our model, in this case the number of subsets in a feasible partition is fixed.

## 2.3 Our Main Results

Here, the main results of this paper are summarised. We start by observing that the GCSG problem is NP–complete on general graphs, even for edge sum valuation functions (Section 3). Alongside the hardness result, we show that a general instance with $|N| = n$ nodes and $|E| = e$ edges can be solved in time $O\left(n^2 \binom{e+n}{n}\right)$ (see Theorem 3).

In order to improve the time required for solving the problem, we make use of tree decompositions. We show that for a graph of $n$ nodes with a tree decomposition of width $w$, the GCSG problem is $O(w^{w+O(1)}n)$. This allows us to derive an upper bound on the computational complexity of GCSG for certain classes of graphs, namely graphs of bounded treewidth, graphs with separator theorems and, in particular, planar graphs and minor–free graphs. We also show that the subclass of edge sum GCSG problems is NP–hard over planar graphs and hence, all $K_k$ minor–free graphs for $k \geq 5$ (see Section 5.1).

Planar graphs are an exceptional family where each graph can be drawn in the plane without any edge crossing. Apart from some interesting mathematical properties such as, for example, 4–colourability and 3–path separability, planar graphs have many practical applications, including design problems for circuits, subways and utility lines. If a network has crossing connections, it usually means that the edges must be run at different heights. While this is not a big issue for electrical wires, it would create extra expenses for some other types of lines—e.g., burying one subway tunnel under another (and therefore deeper than one would normally need). Circuits, in particular, are easier to manufacture if their connections live on fewer layers. Importantly, one may determine a graph's planarity using the so called "forbidden minor" characterisation, by which a graph is planar if and only if it does not contain the complete graph $K_5$ nor the complete bipartite graph $K_{3,3}$ as a minor (Wagner, 1937).[1] Remarkably, such forbidden minor characterisations exist for several graph families that vary in the nature of what is forbidden, and have been utilised in combinatorial algorithms, often for identifying a structure (Robertson & Seymour, 1983, 1995, 2004). This motivates our particular interest in classes of minor–free graphs.

The next theorem is our main technical result.

**Theorem 1** *A general instance of a graph coalition structure generation problem over a graph $G$ with $n$ nodes and a known tree decomposition of width $w$ can be solved in $O(w^{w+O(1)}n)$ computational steps.*

This gives us the immediate corollary.

**Corollary 1** *For any fixed $w$, the GCSG problem over a graph $G$ with $n$ nodes and maximum treewidth $w$ can be solved in $O(n)$ computational steps.*

The proof of these results is presented in Section 4. Coupled with known results regarding separator theorems this gives the base to the following contributions (see Section 5 for proofs).

**Corollary 2** *For any graph $H$ with $k$ vertices, an instance of the graph coalition structure generation problem over an $H$ minor–free graph $G$ with $n$ nodes requires $O(n^{\gamma\sqrt{n}+O(1)})$ computation steps for $\gamma = 0.5k\sqrt{k}/(1 - \sqrt{2/3})$.*

**Corollary 3** *A general instance of a graph coalition structure generation problem over a planar graph $G$ with $n$ nodes can be solved in $O(n^{\gamma\sqrt{n}+O(1)})$ computation steps, for $\gamma = \sqrt{2}/(1 - \sqrt{2/3})$.*

---

1. This characterisation by Wagner's theorem is closely related (but not equivalent) to Kuratowski's theorem, which states that a graph is planar if and only if it does not contain as a subgraph a subdivision of $K_5$ or $K_{3,3}$ (Kuratowski, 1930).

However, for planar graphs we also prove the following hardness result.

**Theorem 2** *The class of edge sum graph coalition structure generation problems over planar graphs is NP–complete. Moreover, a 3-SAT problem with $m$ clauses can be represented by a GCSG problem over a planar graph with $O(m^2)$ nodes.*

Note that Theorem 2 holds for all $K_k$ minor–free graphs where $k \geq 5$, as planar graphs are a special case. This means we should expect it to take time exponential in $\sqrt{n}$ to solve a GCSG problem over such graphs of size $n$. This suggests that the methods given in Corollaries 2 and 3, which solve these problems in time exponential in $\log(n)\sqrt{n}$, are close to the best possible.

Against this background, the main contribution of our work is that it shows significant improvement in complexity of exact algorithms for a general class of coalition structure generation problems characterised by a single assumption of the independence of disconnected members on the valuation functions. In particular, our results are especially valuable for graphs for which a tree decomposition of (low) width can be assessed.

The remaining sections describe our main results and techniques in more detail and contain the proofs.

## 3. General Graphs

In this section, we examine the complexity of coalition structure generation over general graphs. As a first step, we make a technical observation showing that without loss of generality the problem can be restricted to a subset of coalition structures as follows.

**Definition 3** *For a graph $G = (N, E)$, a coalition structure $\mathcal{C}$ over $N$ is connected if the induced subgraph of $G$ over $C$ is connected for all $C \in \mathcal{C}$.*

Lemma 1 will then imply that the GCSG problem is equivalent to maximising the same objective function over all connected coalition structures as in Definition 3. We note that the lemma follows directly from Definition 1 of the IDM property and provide the full proof in the appendix.

**Lemma 1** *Given a graph $G = (N, E)$ and a coalition valuation function $v(\cdot)$ with the IDM property, for any $A, B \subseteq N$ if there are no edges in $G$ between $A \setminus B$ and $B \setminus A$, then*

$$v(A) - v(A \cap B) = v(A \cup B) - v(B).$$

Note, under Definition 1, if $v(\cdot)$ is IDM and we have two coalitions $B$ and $C$ which are disconnected, then by Lemma 1, $v(B \cup C) = v(B) + v(C)$. So, for any coalition $C$, its value $v(C)$ is equal to the sum of $v(\cdot)$ over all its connected components. We can deduce that, for any coalition structure $\mathcal{C}$ there exists a coalition structure $\mathcal{D}$ such that $v(\mathcal{C}) = v(\mathcal{D})$ and all coalitions in $\mathcal{D}$ are connected subgraphs. Thus, without loss of generality, we can restrict our attention to connected coalition structures. Moreover, if $G$ is not a connected graph, then we can solve any coalition structure problem over $G$ with an IDM coalition valuation function by finding the optimal coalition structure over each connected component of $G$ and combining the results. The operation of testing connectivity and finding connected components is computationally tractable in polynomial time (Hopcroft & Tarjan, 1973), and so, without loss of generality, we restrict our attention to connected graphs $G$.

For a (connected) graph $G = (N, E)$ with a set of nodes $N$ and a set of edges $E$, we denote $|N| = n$ and $|E| = e$. Next, we present a simple algorithm for constructing optimal coalition structures over $N$, which is based on the following observation. Note that every connected coalition structure over $N$ can be expressed as the connected components of some subgraph $G' = (N, E')$ of $G$, where $E' \subseteq E$. Moreover, each connected component has a spanning subtree, so we can restrict our attention to *acyclic* subgraphs of $G$. Given this, Algorithm 1 below runs through all acyclic subgraphs of $G$ and their connected components, that correspond to connected coalition structures over the set of nodes $N$. We would like to remark that the order in which the subgraphs of $G$ are checked, has no effect on the outcome, and so can be chosen arbitrarily. Thus, w.l.o.g., we initialise the procedure with a coalition structure $\mathcal{C} = (\{n_1\}, \ldots, \{n_n\})$ that corresponds to connected components of subgraph $G' = (N, \emptyset)$ of $G$.

---

**Algorithm 1** An algorithm for coalition structure generation over general graphs.

---

1:  INPUT: a connected undirected graph $G = (N, E)$;
2:         an IDM coalition valuation function $v : \mathcal{P}(N) \to \mathbb{R}$
3:  OUTPUT: an optimal connected coalition structure over $N$ w.r.t. $v$
4:  $\mathcal{C} \leftarrow (\{n_1\}, \ldots, \{n_n\})$
5:  **for** all $E' \subseteq E$ such that $G' = (N, E')$ is acyclic
6:      find $\mathcal{C}(G') = (\{C_1\}, \ldots, \{C'_k\})$—the collection of all connected components of $G'$
7:      **if** $v\left(\mathcal{C}(G')\right) = \sum_{i=1}^{k'} v(C_i) > v(\mathcal{C})$ **then**
8:        $\mathcal{C} \leftarrow \mathcal{C}(G')$
9:      **end if**
10: **end for**

---

We show the following.

**Theorem 3** *Algorithm 1 solves a general instance of a GCSG problem in $O\left(n^2 \binom{e+n}{n}\right)$ steps, using $O(n \log n)$ sized memory.*

**Proof :** An acyclic subgraph $G' = (N, E')$ of $G$, where $E' \subseteq E$, has at most $n-1$ edges, and so there are at most $\sum_{k=0}^{n-1} \binom{e}{k}$ such subgraphs. Since $\binom{a}{b} + \binom{a}{b-1} = \binom{a+1}{b}$ and $\binom{a}{b} \leq \binom{a+1}{b}$, this sum is bounded by $\binom{e+n}{n}$. Now, it takes at most $O(n^2)$ steps to determine the connected components of a subgraph, and, thus, there are at most $O\left(n^2 \binom{e+n}{n}\right)$ steps needed to check each coalition structure. Finally, it takes at most $O(n \log n)$ sized memory to store each coalition as it is checked. □

Coupled with Corollary 2.3 in the paper by P. Stănică (2001), Theorem 3 implies the following result for sparse graphs.

**Corollary 4** *For sparse graphs with $e = cn$ edges, where $c$ is a constant, the GCSG problem is $O\left(n^{3/2} y^n\right)$ with a constant $y = \frac{(c+1)^{c+1}}{c^c}$.*

This is an easy and not particularly promising result, as it may be exponential in $n \log(n)$ and is exponential in $n$ even for sparse graphs. Indeed, the class of graph coalition structure generation problems is NP–hard: it contains the subclass of GCSG problems over complete graphs, which is equivalent to the NP–complete class of standard coalition structure generation problems over node sets. Importantly, the problem remains hard even for simple coalition valuation functions, such as the correlation function (Bansal et al., 2003). We note that the same holds for the edge sum function as well: this result can be seen as a corollary of Theorem 2 showing the hardness of the edge sum GCSG over planar graphs.

## 4. Tree Decompositions

We now consider solving the GCSG problem over graphs with known tree decompositions. Specifically, we prove our main technical result (Theorem 1) giving a general bound for the GCSG on these graphs, and then derive Corollary 1 regarding graphs with bounded treewidth. The proof follows by recursively calculating the potential marginal contributions to total coalition structure valuation for each branch of a tree decomposition (see Algorithm 2). To build the intuition, we first derive two technical lemmas. For brevity of exposition, their proofs are presented in the Appendix.

**Lemma 2** *Let $G = (N, E)$ be a graph with a tree decomposition $(X, T)$, where $X = \{X_1, \ldots, X_m\}$ for $m \leq n = |N|$ and $T$ is a tree over $X$. Suppose further that the $X_i$ are numbered in order of shortest distance in $T$ from $X_1$, where $X_1$ may be chosen arbitrarily. Then, for any $C \subseteq N$,*

$$v(C) = \sum_{i=1}^{m} v(C \cap X_i) - v\big(C \cap X_i \cap \bigcup_{j<i} X_j\big).$$

Lemma 2 above will allow us to calculate the value of a total coalition structure from local structures defined on branches of a tree decomposition. We now discuss how to construct such a total structure from the local ones. We need the following notation.

For any graph $G = (N, E)$, for any $P, Q \subseteq N$, if $\mathcal{P}$ is a coalition structure over $P$ and $\mathcal{Q}$ is a coalition structure over $Q$, then we define

$$U(\mathcal{P}, \mathcal{Q}) = \{A \in \mathcal{P} : A \subseteq (P \backslash Q)\} \cup \{B \in \mathcal{Q} : B \subseteq (Q \backslash P)\} \cup \{A \cup B : A \in \mathcal{P}, B \in \mathcal{Q}, A \cap B \neq \emptyset\}.$$

That is, $U(\mathcal{P}, \mathcal{Q})$ is a collection of subsets of $P \cup Q$ that agrees with $\mathcal{P}$ over $P \setminus Q$ and with $\mathcal{Q}$ over $Q \setminus P$, and contains all pairwise unions of subsets $A \in \mathcal{P}$ and $B \in \mathcal{Q}$ with non-empty intersections. Note that $U(\mathcal{P}, \mathcal{Q})$ is not necessarily a coalition structure over $P \cup Q$, as the union coalitions $A \cup B$, $A \in \mathcal{P}$, $B \in \mathcal{Q}$, need not be disjoint.

Furthermore, for a graph $G = (N, E)$ and a coalition structure $\mathcal{P}$ over some subset of nodes $P \subseteq N$, for any further subset $P' \subseteq P$ we will denote by $\mathcal{P}(P')$ a coalition structure over $P'$ defined as follows:

$$\mathcal{P}(P') = \{C \cap P' : C \in \mathcal{P}\}.$$

That is, for any $x, y \in P' \subseteq P$, they belong to the same coalition in $\mathcal{P}(P')$ if and only if they belong to the same coalition in $\mathcal{P}$.

For illustration, consider the following example. Let $N = \{1, 2, 3, 4, 5\}$, take two subsets $P = \{1, 2, 3\}$ and $Q = \{3, 4, 5\}$ of $N$, and define coalition structures $\mathcal{P} = \{\{1\}, \{2, 3\}\}$ and $\mathcal{Q} = \{\{3, 4\}, \{5\}\}$ over $P$ and $Q$, respectively. Note that $\{1\} \in \mathcal{P}$ is a subset of $P \setminus Q$, $\{5\} \in \mathcal{Q}$ is a subset of $Q \setminus P$, and $(\{2,3\} \in \mathcal{P}) \cap (\{3,4\} \in \mathcal{Q}) = \{3\}$. Then, $U(\mathcal{P}, \mathcal{Q}) = \{\{1\}, \{5\}, \{2,3\} \cup \{3,4\}\} = \{\{1\}, \{5\}, \{2,3,4\}\}$. Now, let $P' = \{1,2\} \subseteq P$ and $Q = \{4,5\} \subseteq Q$. Then, $\mathcal{P}(P') = \{\{1\} \cap \{1,2\}, \{2,3\} \cap \{1,2\}\} = \{\{1\}, \{2\}\}$ and $\mathcal{Q}(Q') = \{\{3,4\} \cap \{4,5\}, \{5\} \cap \{4,5\}\} = \{\{4\}, \{5\}\}$.

**Lemma 3** *For any graph $G = (N, E)$, for any $P, Q \subseteq N$, if $\mathcal{P}$ is a coalition structure over $P$ and $\mathcal{Q}$ is a coalition structure over $Q$, and if $\mathcal{P}(P \cap Q) = \mathcal{Q}(P \cap Q)$, then $\mathcal{E} = U(\mathcal{P}, \mathcal{Q})$ is a coalition structure over $P \cup Q$ and for any $P' \subseteq P$, and $Q' \subseteq Q$, $\mathcal{E}(P) = \mathcal{P}(P')$ and $\mathcal{E}(Q) = \mathcal{Q}(Q')$.*

We are now ready to prove Theorem 1. To this end, below we present Algorithm 2 that, given a graph with a known tree decomposition, finds best coalition structure over the node set by recursively calculating the potential marginal contributions to total coalition structure valuation for each branch of a given tree decomposition. Lemma 4 below proves its validity and computational bounds.

---

**Algorithm 2** An algorithm for coalition structure generation over graphs with known tree decompositions.

---

1:    INPUT: a connected undirected graph $G = (N, E)$;
2:        a tree decomposition $(X, T)$ of $G$, where $X = \{X_1, \ldots, X_m\}$ for $m \leq n$,
3:        $T$ is a tree over $X$, and $1 \leq i < j \leq m \Leftrightarrow d_T(X_i, X_1) \leq d_T(X_j, X_1)$, where
4:        for any $1 \leq i \leq m$, $d_T(X_i, X_1)$ is the distance of $X_i$ from $X_1$
5:        an IDM coalition valuation function $v : \mathcal{P}(N) \to \mathbb{R}$
6:    OUTPUT: an optimal connected coalition structure over $N$ w.r.t. $v$
7:    **for all** $1 \leq i \leq m$
8:        $Y_i \leftarrow X_i \setminus \cup_{j<i} X_j$
9:        $Z_i \leftarrow X_i \setminus Y_i$
10:      $D_i \leftarrow \{j > i : (X_i, X_j) \in T\}$
11:   **for** $i = m, m-1, \ldots, 1$
12:      **for all** $\mathcal{C}$—coalition structures over $Z_i$
13:        $v_i(\mathcal{C}) \leftarrow \max_{\mathcal{E}} \sum_{C \in \mathcal{E}} \big( v(C) - v(C \setminus Y_i) \big) + \sum_{j \in D_i} v_j(\mathcal{E}(Z_j))$,
14:        where $\mathcal{E}$ are coalition structures over $X_i$ such that $\mathcal{E}(Y_i) = \mathcal{C}$
15:      **end for**
16:   $\mathcal{C}_0 \leftarrow \arg\max_{\mathcal{C}} v_1(\mathcal{C})$ where $\mathcal{C}$ are colition structures over $Z_1$
17:   **for** $k = 1, \ldots, m$
18:      $\mathcal{C}_k \leftarrow U(\mathcal{C}_{k-1}, \mathcal{E}_k)$,
19:      where $\mathcal{E}_k$ is any coalition structure over $X_k$ such that $\mathcal{E}_k(Z_k) = \mathcal{C}_{k-1}(Z_k)$
20:      and $v_k(\mathcal{C}_{k-1}(Z_k)) = \sum_{C \in \mathcal{E}_k} \big( v(C) - v(C \setminus Y_k) \big) + \sum_{j \in D_k} v_j(\mathcal{E}_k(Z_j))$
21:   **end for**
22:   output $\mathcal{C}_m$

---

**Lemma 4** *Algorithm 2 solves a general instance of a graph coalition structure generation problem over a graph $G$ with $n$ nodes and a known tree decomposition of width $w$ in $O(w^{w+O(1)} n)$ computational steps.*

**Proof :**   We are given a graph $G = (N, E)$ with a tree decomposition $(X, T)$, where $X = \{X_1, \ldots, X_m\}$ for $m \leq n = |N|$ and $T$ is a tree over $X$. Suppose for some $w$, $|X_i| < w$ for all $i$. We assume without loss of generality that the $X_i$ are numbered in order of shortest distance in $T$ from $X_1$, where $X_1$ may be chosen arbitrarily. Thus, for each $i > 1$, $X_i$ has exactly one link in $T$ that connects to an $X_j$ with $j < i$. For each $i$ we define $Y_i$ to be $X_i \setminus \cup_{j < i} X_j$ and $Z_i$ to be $X_i \setminus Y_i$. Note, for each $i > 1$ there exists a single $j < i$ such that $Z_i \subseteq X_j$, and hence $Z_i = (X_j \cap X_i)$. Since every node must be in at least one $X_i$, we have that the union of the $Y_i$ is $N$. Finally, for each $i$, $D_i$ is the set of $j > i$ such that $(X_i, X_j)$ is an edge in $T$.

Now, for each $i = m, m - 1, \ldots, 1$, the algorithm recursively define functions $v_i(\cdot)$ which give real values for each coalition structure over $Z_i$. For $\mathcal{C}$, a coalition structure over $Z_i$, we let $v_i(\mathcal{C})$ be the maximum of

$$\sum_{C \in \mathcal{E}} \big( v(C) - v(C \setminus Y_i) \big) + \sum_{j \in D_i} v_j(\mathcal{E}(Z_j)),$$

over all coalition structures $\mathcal{E}$ over $X_i$ such that $\mathcal{E}(Y_i) = \mathcal{C}$. Note for any $j \in D_i$, $Z_j = (X_i \cap X_j)$, and hence, for any coalition structure $\mathcal{E}$ over $X_i$, $\mathcal{E}(Z_j)$ forms a coalition structure over $Z_j$.

Now, suppose $\mathcal{C}$ is a coalition structure over $G$. We will show that $v(\mathcal{C}) \leq v_1(\mathcal{C}(Z_1))$. We do this by showing inductively that, for all $k \geq 1$,

$$v_1(\mathcal{C}(Z_1)) \geq \sum_{i=1}^{k} \sum_{C \in \mathcal{C}(X_i)} \big( v(C) - v(C \setminus Y_i) \big) + \sum_{j \in D_i : j > k} v_j(\mathcal{C}(Z_j)). \tag{1}$$

For $k = 1$ this follows from the definition of $v_1(\cdot)$, as $\mathcal{C}(X_1)$ is a coalition structure over $X_1$. Now it is sufficient to show that the right hand side of (1) does not increase as $k$ increases. For general $k$ the change in the right hand side of (1) from the preceeding iteration is

$$\sum_{C \in \mathcal{C}(X_k)} \big( v(C) - v(C \setminus Y_k) \big) + \sum_{j \in D_k} v_j(\mathcal{C}(Z_j)) - v_k(\mathcal{C}(Z_k)).$$

It follows from the definition of $v_k(\cdot)$ that this value is non-positive, as $(\mathcal{C}(X_k))$ is a coalition structure over $X_k$. Hence, the inductive proof is complete. Thus, we have shown that

$$v_1(\mathcal{C}(Z_1)) \geq \sum_{i=1}^{m} \sum_{C \in \mathcal{C}(X_i)} \big( v(C) - v(C \setminus Y_i) \big) = v(\mathcal{C}),$$

by Lemma 2. So, the maximum of $v_1(\mathcal{E})$ for coalition structures $\mathcal{E}$ over $Z_1$ is greater than or equal to the maximum value of $v(\mathcal{C})$ over all coalition structures $\mathcal{C}$ over $G$.

Now, let $\mathcal{C}_0$ be a coalition structure over $Z_1$ that maximises $v_1(\mathcal{C})$. The algorithm recursively defines coalition structures $\mathcal{C}_1, \mathcal{C}_2, \ldots \mathcal{C}_m$ by setting, for all $1 < k \leq m$, $\mathcal{C}_k = U(\mathcal{C}_{k-1}, \mathcal{E}_k)$, where $\mathcal{E}_k$ is any coalition structure over $X_k$ such that $\mathcal{E}_k(Z_k) = \mathcal{C}_{k-1}(Z_k)$ and

$$v_k(\mathcal{C}_{k-1}(Z_k)) = \sum_{C \in \mathcal{E}_k} \big( v(C) - v(C \setminus Y_k) \big) + \sum_{j \in D_k} v_j(\mathcal{E}_k(Z_j)).$$

We now want to show that

$$v_1(\mathcal{C}(Z_1)) = \sum_{i=1}^{k} \sum_{C \in \mathcal{C}_k(X_i)} (v(C) - v(C \setminus Y_i)) + \sum_{j \in D_i : j > k} v_j(\mathcal{C}_k(Z_j)). \qquad (2)$$

Again, we use induction. For $k = 1$, this follows from the definition of $v_1(\cdot)$, by noting that since $\mathcal{C}_1 = U(\mathcal{C}_0, \mathcal{E}_1)$, Lemma 3 implies that $\mathcal{C}_1(X_1) = \mathcal{E}_1(X_1)$, but since both are coalition structures over $X_1$, we must have $\mathcal{C}_1 = \mathcal{E}_1$.

Now, for general $k$, since $\mathcal{C}_k = U(\mathcal{C}_{k-1}, \mathcal{E}_k)$, we must have, for all $i < k$, $\mathcal{C}_k(X_i) = \mathcal{C}_{k-1}(X_i)$, and for all $j \in D_i$ such that $j \geq k$, since $Z_j \subseteq X_i$, we have, $\mathcal{C}_k(Z_j) = \mathcal{C}_{k-1}(Z_j)$. Thus, the change in the right hand side of (2) from the previous increment is equal to

$$\sum_{C \in \mathcal{C}_k(X_k)} (v(C) - v(C \setminus Y_i)) + \sum_{j \in D_k} v_j(\mathcal{C}_k(Z_j)) - v_k(\mathcal{C}_k(Z_k))$$
$$= \sum_{C \in \mathcal{E}_k(X_k)} (v(C) - v(C \setminus Y_i)) + \sum_{j \in D_k} v_j(\mathcal{E}_k(Z_j)) - v_k(\mathcal{C}_{k-1}(Z_k)) = 0,$$

by the definition of $\mathcal{C}_k$ and $\mathcal{E}_k$. This completes this inductive proof.

So we have shown that

$$v_1(\mathcal{C}(Z_1)) = \sum_{i=1}^{m} \sum_{C \in \mathcal{C}_m(X_i)} (v(C) - v(C \setminus Y_i)) = v(\mathcal{C}_m).$$

Since $v_1(\mathcal{C}(Z_1))$ is an upper bound for $v(\cdot)$ over all coalition structures on $N$, we must have that $\mathcal{C}_m$ is a solution to our coalition valuation problem.

Finally, in order to solve the coalition problem, all that needs to be done is to fully calculate $v_k(\cdot)$ for each $k$ from $m$ down to 1, recording corresponding optimal coalition structures for each value, and then optimise $v_1(\cdot)$. To do this, for each $k$, we can go through each coalition structure $\mathcal{E}$ over $X_k$, and then calculate

$$\sum_{C \in \mathcal{E}} \big( v(C) - v(C \setminus Y_i) \big) + \sum_{j \in D_i} v_j(\mathcal{E}(Z_j)).$$

If this is greater than the currently held value for $v_k(\mathcal{E}(Z_k))$, then replace that value and also record $\mathcal{E}$. This requires polynomial (in $w$) calculations for each possible coalition structure over each node $X_k$, which gives $O(w^{w+O(1)})$ calculations for each $X_k$ and thus $O(w^{w+O(1)}n)$ calculations in total. □

Theorem 1 follows immediately from Algorithm 2 and Lemma 4. Now, given any $w$, for the class of graphs of maximum treewidth $w$, a tree decomposition with width at most $w$ may be found in linear time (Bern, Lawlerand, & Wong, 1987). Given this, Corollary 1 below is directly implied by Theorem 1.

**Corollary 1** *For any fixed $w$, the GCSG problem over a graph $G$ with $n$ nodes and maximum treewidth $w$ can be solved in $O(n)$ computational steps.*

If we set $w = 1$, then this result applies to acyclic graphs, and is related to results of De-mange (2004) regarding coalition structure generation over trees. However, Demange (2004) does not make the IMD assumption. Their resulting algorithm is more complex than ours and has potentially exponential running time. This is to be expected, as without the in-dependence of disconnected members, the coalition structure generation problem over star networks is necessarily exponential.

Note, if we set $w = 2$, then the class of graphs under consideration becomes the class of $K_4$ minor–free graphs. Likewise, the class of graphs of treewidth 1 may be characterised as $K_3$ minor–free. These results are in sharp contrast to Theorem 2 which shows NP-completeness for the edge sum GCSG problem over planar graphs, which are a subset of the class of $K_5$ minor–free graphs. We give a proof of Theorem 2 in the next section.

## 5. Separator Theorems

In this section, we prove computational bounds for the GCSG problem over minor–free and planar graphs. These graphs are guaranteed to contain vertex separators, as formalised by Definition 4 below. Intuitively, this means that graphs in the corresponding class can be split into smaller pieces by removing a small number of vertices. In general,

**Definition 4** *A class of graphs $\mathcal{G}$ satisfies an $f(n)$-separator theorem with constant $\alpha < 1$ if for all $G = (N, E) \in \mathcal{G}$ with $|N| = n$ there exists a subset $S \subseteq N$ such that $|S| \leq f(n)$ and $N \setminus S = A \cup B$ for disjoint $A$ and $B$ where, $|A| \leq \alpha n$, $|B| \leq \alpha n$, and there exists no $x \in A$ and $y \in B$ such that $(x, y) \in E$.*

To illustrate this, consider for example a grid graph $G$ with $r$ rows and $c$ columns, where $n = rc$ is the number of nodes. If $r$ is odd, then there is a single central row, and otherwise, there are two rows equally close to the center; similarly, if $c$ is odd, then there is a single central column, and otherwise, there are two columns equally close to the center. Let a node subset $S$ be any of these central rows or columns. Removing $S$ from the graph will divide it into two smaller disjoint components, $A$ and $B$, each of which has at most $n/2$ vertices. If $r \leq c$, then a central column defines a separator $S$ with $r \leq \sqrt{n}$ vertices, and similarly, if $c \leq r$, then a central row is a separator with at most $\sqrt{n}$ vertices. Thus, any grid graph has a separator $S$ of size at most $\sqrt{n}$, the removal of which splits the graph into two connected components, each of size at most $n/2$. That is, the class of grid graphs satisfy a $\sqrt{n}$-separator theorem with constant $\alpha = 1/2$.

We now use Theorem 1 to derive Algorithm 3 and Lemma 5, which provide us with a general result for classes of graphs that satisfy separator theorems. We will then apply this result to the classes of minor–free and planar graphs, coupled with their corresponding separator theorems, to obtain computational bounds on coalition structure generation over these graphs.

Suppose we have a class of graphs $\mathcal{G}$ that is closed under taking subgraphs and satisfies an $f(n)$-separator theorem with constant $\alpha < 1$, where $f(n) = \beta n^c$ for some constants $\beta, c$, and there exists an algorithm to find such a separator for any $G \in \mathcal{G}$ with $n$ nodes in polynomial time. Given this, for any such graph $G \in \mathcal{G}$, Algorithm 3 below finds a tree decomposition with treewidth $w \leq \beta n^c / (1 - \alpha^c)$ in polynomial time. Our procedure is based on the proof of Theorem 20 in the work of Bodlaender (1998), which states that for any such class of graphs $\mathcal{G}$, the treewidth of any $G \in \mathcal{G}$ with $n$ nodes is $O(f(n))$. We then apply Algorithm 2 to solve the GCSG problem for $G$ with this tree decomposition in $O(w^{w+O(1)}n)$ computational steps, which finally provides us with a computational bound of $O(n^{\frac{\beta c}{1-\alpha^c} n^c + O(1)})$ time, as stated in Lemma 5 below.

---

**Algorithm 3** An algorithm for coalition structure generation over graphs with separator theorems.

1:    INPUT: a graph $G = (N, E) \in \mathcal{G}$; an IDM coalition valuation function $v : \mathcal{P}(N) \to \mathbb{R}$
2:    OUTPUT: an optimal connected coalition structure over $N$ w.r.t. $v$
3:    **if** $n = 1$ **then**
4:        $\tilde{X} \leftarrow (\{x\})$, where $x \in N$ is the only node of $G$
5:        $T' \leftarrow G$
6:    **otherwise**
7:        find $S$, a $\beta n^c$ separator of $G$ with $N \setminus S = A \cup B$ where $|A| \leq \alpha n$ and $|B| \leq \alpha n$
8:        find tree decompositions $(X^A, T^A)$ of $A$ and $(X^B, T^B)$ of $B$, both of width $\leq \frac{\beta \alpha^c n^c}{1-\alpha^c}$
9:        $T' \leftarrow T^A \cup T^B \cup \{e'\}$ where $e' = \{x, y\} \in E$, $x \in A$, $y \in B$
10:      $\tilde{X}^A \leftarrow \{X \cup S : X \in X^A\}$
11:      $\tilde{X}^B \leftarrow \{X \cup S : X \in X^B\}$
12:      $\tilde{X} \leftarrow \tilde{X}^A \cup \tilde{X}^B$
13:    **end if**
14:    apply Algorithm 2 to $G$ with tree decomposition $(\tilde{X}, T')$

---

**Lemma 5** *Let $\mathcal{G}$ be a class of graphs that is closed under taking subgraphs and satisfies an $f(n)$-separator theorem with constant $\alpha < 1$, where $f(n) = \beta n^c$ for some constants $\beta, c$, and there exists an algorithm to find such a separator for any $G \in \mathcal{G}$ with $n$ nodes in polynomial time. Then, Algorithm 3 solves a GCSG problem over a graph $G \in \mathcal{G}$ with $n$ nodes in $O(n^{\gamma n^c + O(1)})$ time, for*

$$\gamma = \frac{\beta c}{1 - \alpha^c}.$$

**Proof :** Suppose we have a class of graphs $\mathcal{G}$ satisfying the statement of this lemma. There must exist constants $K$ and $d > -\log_\alpha(2)$ such that for any $G \in \mathcal{G}$ with $n$ nodes, we can find a $\beta n^c$ separator, with constant $\alpha$, in $K n^d$ computational steps. Our proof then proceeds by showing that, for any $G \in \mathcal{G}$ with $n$ nodes, Algorithm 3 (steps 3–13) finds a tree decomposition with width less than or equal to $\beta n^c / (1 - \alpha^c)$ in at most $K n^d / (1 - 2\alpha^d)$ computational steps. We prove the result by induction.

For $n = 1$ no computational steps are required as $G$ is already in tree form. For the $n$th inductive step, suppose we have a $G = (N, E)$ in $\mathcal{G}$ with $|N| = n$. In $K n^d$ computational steps we can find $S$, a $\beta n^c$ separator of $G$ with $N \setminus S = A \cup B$ where $|A| \leq \alpha n$ and $|B| \leq \alpha n$. By the inductive hypothesis, we can apply steps 3–13 of Algorithm 3 to find tree decompositions $(X^A, T^A)$ and $(X^B, T^B)$ of the subgraphs $A$ and $B$ respectively, taking a total time of $2 K \alpha^d n^d / (1 - 2\alpha^d)$, where $(X^A, T^A)$ and $(X^B, T^B)$ both have maximal width $\beta \alpha^c n^c / (1 - \alpha^c)$. Now, let $\tilde{X}^A = \{X \cup S : X \in X^A\}$, let $\tilde{X}^B = \{X \cup S : X \in X^B\}$, and let $T'$ be any tree formed by connecting $T^A$ and $T^B$ by a single edge. Then, we claim $(\tilde{X}^A \cup \tilde{X}^B, T')$ is a tree decomposition of $G$. For any $a \in A \setminus S$, the set of elements of $X^A$ that $a$ appears in, forms a subtree of $T^A$, and thus, the set of elements of $\tilde{X}^A \cup \tilde{X}^B$ that $a$ appears in, must form a subtree of $T'$. By symmetry, the same holds for $a \in B \setminus S$. Further, for $a \in S$, $a$ appears in every element of $\tilde{X}^A \cup \tilde{X}^B$. Lastly, for each pair of nodes connected by an edge in $G$, if those nodes both lie inside $A$ or $B$, then they will both be in some element of $\tilde{X}^A$ or $\tilde{X}^B$ respectively, otherwise at least one of those nodes must lie in $S$, and so must be a member of every element of $\tilde{X}^A \cup \tilde{X}^B$. This proves our claim. The tree decomposition $(\tilde{X}^A \cup \tilde{X}^B, T')$ took at most

$$Kn^d + \frac{2K\alpha^d n^d}{1 - 2\alpha^d} = \frac{Kn^d}{1 - 2\alpha^d}$$

computational steps to find and has width at most

$$\beta n^c + \frac{\beta \alpha^c n^c}{1 - \alpha^c} = \frac{\beta n^c}{1 - \alpha^c},$$

as required. This completes our inductive proof.

Thus, for any $G \in \mathcal{G}$ with $n$ nodes, we can find a tree decomposition for $G$ with treewidth at most $\beta n^c / (1 - \alpha^c)$ in polynomial time. We can now apply Algorithm 2, to solve the GCSG problem for a graph $G \in \mathcal{G}$ with $n$ nodes in $O(w^{w+O(1)}n)$ computational steps, where $w = \beta n^c / (1 - \alpha^c)$. However,

$$w^w = \frac{\beta}{1 - \alpha^c} n^{c\beta n^c / (1 - \alpha^c)} = O(n^{\gamma n^c}),$$

and so the statement of the lemma follows. □

This result allows us to obtain computational bounds for the GCSG problem over minor–free and planar graphs as follows.

**Corollary 2** *For any graph $H$ with $k$ vertices, an instance of the graph coalition structure generation problem over an $H$ minor–free graph $G$ with $n$ nodes requires $O(n^{\gamma \sqrt{n} + O(1)})$ computation steps for $\gamma = 0.5k\sqrt{k}/(1 - \sqrt{2/3})$.*

**Proof :** We apply Lemma 5 using the main result in the paper by Alon, Seymour and Thomas (1990) where it was shown that the class of such graphs satisfies a $k\sqrt{kn}$-separator theorem with $\alpha = 2/3$. Thus, we can solve a general instance of the problem in $O(n^{\gamma \sqrt{n} + O(1)})$ for $\gamma = k\sqrt{k}/2(1 - \sqrt{2/3})$, as required. □

It should be noted that Proposition 4.5 of Alon, Seymour and Thomas (1990) gives a bound of $k\sqrt{kn}$ on the treewidth of this class of graphs, but it is not constructive, so cannot be combined with Theorem 1 as this requires a tree decomposition to be available.

For planar graphs, Corollary 3 provides a stronger result.

**Corollary 3** *A general instance of a graph coalition structure generation problem over a planar graph $G$ with $n$ nodes can be solved in $O(n^{\gamma \sqrt{n} + O(1)})$ computation steps, for*

$\gamma = \sqrt{2}/(1 - \sqrt{2/3})$.

**Proof :** We apply Lemma 5 using the main result in the work of Lipton and Tarjan (1979) where it was shown that the class of such graphs satisfies a $2\sqrt{2n}$-separator theorem with $\alpha = 2/3$. Thus, we can solve a general instance of the problem in $O(n^{\gamma\sqrt{n}+O(1)})$ for $\gamma = \sqrt{2}/(1 - \sqrt{2/3})$, as required. $\square$

Recall that the class of planar graphs is equivalent to the class of $K_{3,3}$ and $K_5$ minor–free graphs. For these graphs, Theorem 2 shows that the graph coalition structure generation problem is NP–complete, even for simple, edge sum, coalition valuation functions (the proof of the theorem is presented in 5.1 below). However, as mentioned in the previous section, the GCSG over smaller minor–free instances can be solved in linear time.

## 5.1 Planar Graphs

Here we prove NP-hardness result for planar graphs. Since planar graphs are $K_5$ minor free, the same hardness result must hold for the class $K_k$ minor–free graphs for all $k \geq 5$. The proof proceeds by finding a representation of a general 3-SAT problem as a GCSG problem over a planar graph.

**Theorem 2** *The class of edge sum graph coalition structure generation problems over planar graphs is NP–complete. Moreover, a 3-SAT problem with $m$ clauses can be represented by a GCSG problem over a planar graph with $O(m^2)$ nodes.*

**Proof :** Suppose we have a 3-SAT problem with clauses $C_1, \ldots C_m$. We will construct an edge sum graph coalition structure generation problem over a planar graph of $O(m^2)$ nodes which, when solved, reveals a solution to the 3-SAT problem if one exists. We will use a series of diagrams to define some components from which we can construct an appropriate edge sum graph. Our diagrams will denote edge values using the symbols given in the key in Figure 1.

The first component is given in Figure 2. We will use the symbol in Subfigure 2b to represent three nodes that surround a subgraph with edge values given in Subfigure 2a. If this is a subgraph of an edge sum problem graph, then the contribution these edge values make to the valuation of a coalition structure is at most 2, with equality only if the induced structure over the three outer nodes is as shown in one of Subfigure 2c, Subfigure 2d or Subfigure 2e. If the induced coalition structure over these three nodes is not one of these two structures, then the contribution will be less than 2. We similarly describe two more triangular components in Figures 2, 4 and 5. The planar graph edge sum problem we construct will be created from these components, some of which will be connected by edges with value 1, others of which will overlap, in the sense that they will share nodes. We will have components sharing nodes with each other, but they will not share edges. Moreover, components can only share those nodes that form the triangle which borders the component. If two components share a pair of such nodes, we will represent this symbolically by drawing their symbols as being adjacent to each other along the corresponding side of the triangular symbols. So, the edges of the symbols of components will touch, but this does not mean that those components share an edge within the graph.

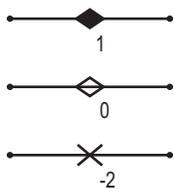

Figure 1: Edge value key.

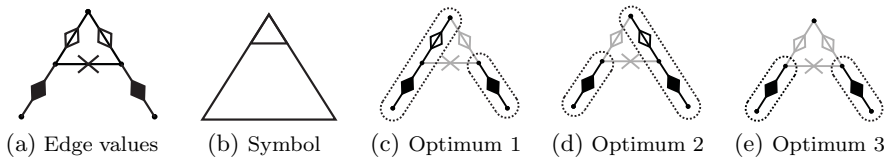

Figure 2: Edge sum problem component.

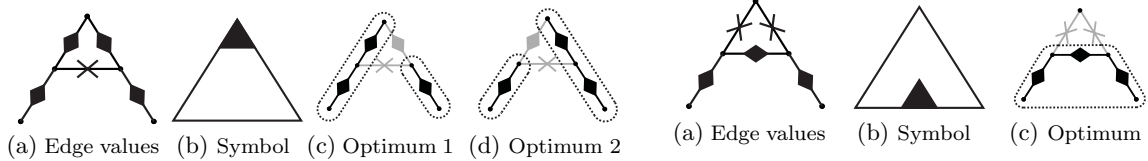

Figure 3: Edge sum problem component.

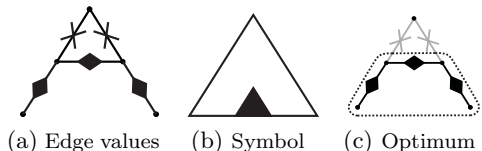

Figure 4: Edge sum problem component.

For a graph consisting of these components, constructed in this way, we will say that a coalition structure is locally optimal if the induced structure over every component is optimal for that component and every connecting edge that is not part of a component lies inside a coalition. For every coalition structure, for each component, the contribution that the edges of that component make to the value of the coalition structure is bounded by the local optimum. Thus, if a coalition structure is locally optimal then it must be optimal. Furthermore, the coalition value of such a coalition structure is straightforward to calculate - simply sum the local optimums of each component and connecting edge. Note, the value obtained by doing this always represents an upper bound on the total valuation of any coalition structure, thus if a locally optimal structure exists, then all optimal coalition structures must be locally optimal. However, it is not guaranteed that a locally optimal structure will exist.

With this in mind, it is now possible to provide some intuition regarding our components. The component in Figure 5 is such that a coalition structure can not be locally optimal unless the three nodes that form its outer triangle either all lie in the same coalition or all in different coalitions. The component in Figure 3 is such that a coalition structure can not be locally optimal unless exactly one of the bottom two nodes is in the same coalition as the top node. The component in Figure 2 is similar to that in Figure 3, except it allows the addition possibility that a locally optimal coalition structure has all three outer nodes in different coalitions. For the component in Figure 4, a coalition structure can only be locally optimal if the bottom two node are in the same coalition, and this coalition does not contain the top node. We will now describe some constructs which are made from the above described components. The first is given in Figure 6. It is such that in any locally optimal coalition structure, nodes X and Y are always in the same coalition and the pair of nodes labelled A lie in the same coalition as each other if and only if the pair of nodes

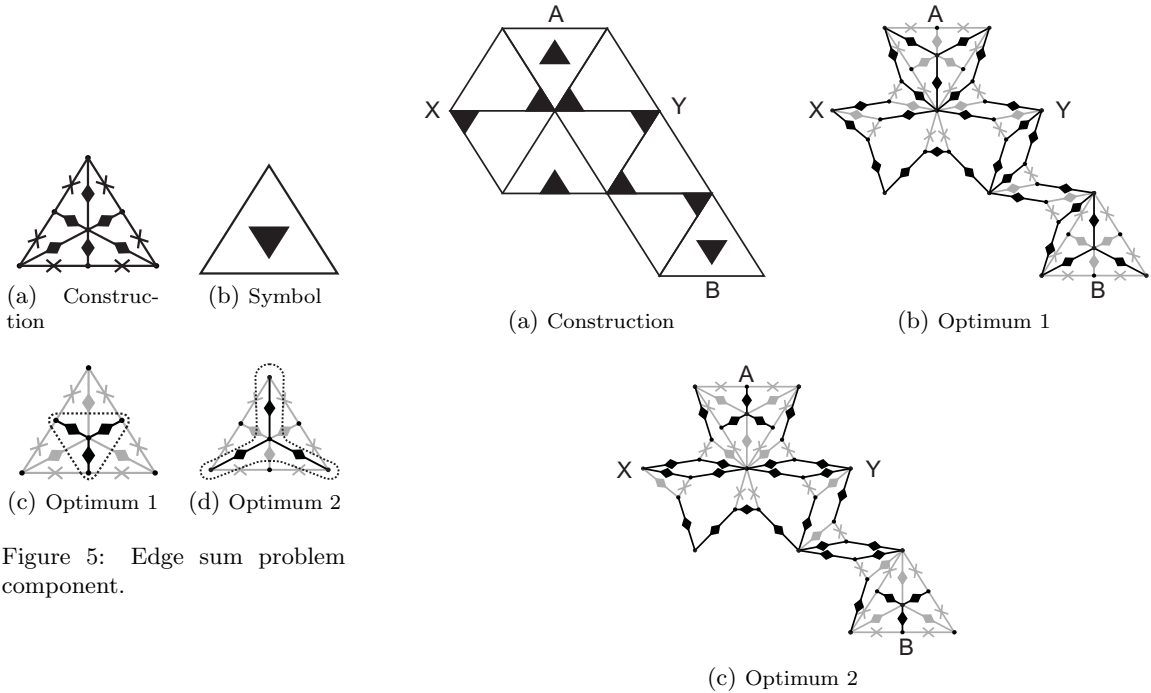

(a) Construction    (b) Symbol

(c) Optimum 1    (d) Optimum 2

Figure 5: Edge sum problem component.

(a) Construction

(b) Optimum 1

(c) Optimum 2

Figure 6: Edge sum problem construct.

labelled B lie in the same coalition as each other. In our reduction of 3-SAT problems, we will be representing logical states by whether or not certain pairs of agents lie in the same coalition in a locally optimal coalition structure. This construct allows us to enforce that two pairs represent the same logical state whilst also allowing a coalition to passes between them in the plane.

The second and third constructs are given in Figures 7 and 8. In the second construct, under a locally optimal coalition structure, if the pair of nodes labelled A are together in the same coalition, then the pair of nodes labelled B are in the same coalition, and similarly for the pair of nodes labelled C. If the pair of nodes labelled A are not in the same coalition, then the pair of nodes labelled B are not in the same coalition, and similarly for the pair of nodes labelled C. Thus, in our representation of a 3-SAT problem, in a locally optimal solution the pairs of nodes labelled A, B and C will always represent the same logical state. The third construct is similar, except that under a locally optimal coalition structure, the state of whether or not the pair of nodes labelled C are in the same coalition as each other is the opposite to the state of the other two pairs of nodes. Thus, in our representation of a 3-SAT problem, in a locally optimal solution the pairs of nodes labelled A and B will represent the same logical state, while C will represent the negation of that state. The last construct is given in Figure 9. It is more complex than the other constructs, so we shall first examine three subgraphs of it. The first part, AX, consists of the subgraph of the three components from the pair of nodes labelled A to the pair of nodes labelled X, the second, BY covers the three components from Y to B and CZ consists of the bottom two components. Note the middle triangle in the diagram with edges X, Y, Z, is not a component, it is merely empty space. Subfigures 9b–9h show all the locally optimal coalition structures for each of these three parts (with only the outer nodes for each component being shown). Since the construct is the union of these three parts, if a coalition structure is locally optimal over each of these subgraphs, then it is locally optimal over the whole construct. However, not every combination of these local optimums is possible. For, if a coalition structure induces Subfigure 9b over AX and Subfigure 9d over BY then the three node in triangle XYZ must lie in the same coalition, and it is not possible for that coalition structure to induce Subfigure 9f. For a locally optimal coalition structure, it cannot be true that each node in A, B and C lies in a different coalition than the node it is paired with. Suppose we think of a pair of nodes as representing a false state if they lie in the same coalition and a true state if they are in different coalitions. Then, for a locally optimal coalition structure over this construct, at least one of A, B and C must represent a true state. It is straightforward to check that there exist locally optimal coalition structures over this construct that induce every possible combination of states besides that where A, B and C are all represent falsehood. Thus, this construct enforces a logical OR within our 3-SAT solution representation. We construct our edge sum problem to represent a general 3-SAT problem as follows. We create a copy of the construct in Figure 9 for each clause of the problem. The three pairs labelled $A, B, C$ are identified with the three literals in the corresponding clause. We identify a coalition structure over these constructs with a set of logical values for the literals in the clauses by saying that the literal associated with a pair of node is set as true if and only if those nodes lie inside a single coalition. For each variable we create a path of copies of the constructs in Figures 7 and 8, where the pair of nodes labelled B for one component are shared and labelled A in the following component. This path should include a copy of the construct in Figure 7 for each literal representation of the variable, and a copy of the construct in Figure 8 for each literal representation of the variable's negation. We then connect each pair of nodes that represents a literal representation of the variable or its negation to the pair of nodes labelled C on it's corresponding construct in the path, using a parallel pair of connecting edges, each of value 1. This ensures that any locally optimal coalition structure has to assign consistent logical values to literal representations of each variable and its negative. To ensure that the resulting graph is planar, we can replace any two parallel pairs of connecting edges which cross over each other with two copies of the construct in Figure 6. For, if there are two copies of the construct in Figure 6 where the first copy shares the nodes labelled B with the nodes labelled A in the second copy, then, under a locally optimal coalition structure, the logical value represented by the nodes labelled A in the first construct will equal the logical value represented by the nodes labelled B in the second construct. Furthermore the logical value represented by the nodes labelled X in the two constructs will equal the logical value of the nodes labelled Y in the two constructs. This allows logical values to "pass each other" in the plane.

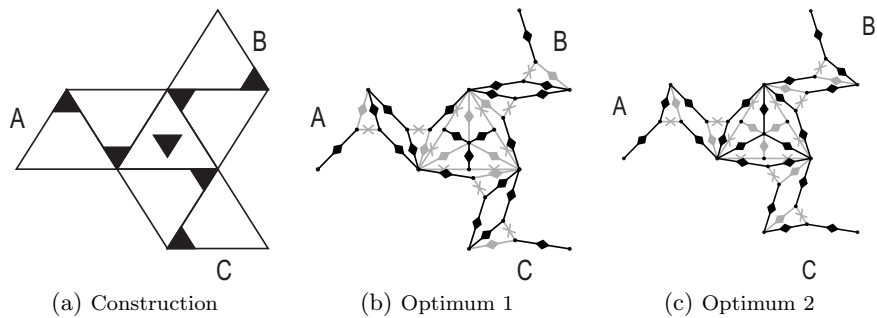

| (a) Construction | (b) Optimum 1 | (c) Optimum 2 |

Figure 7: Edge sum construct.

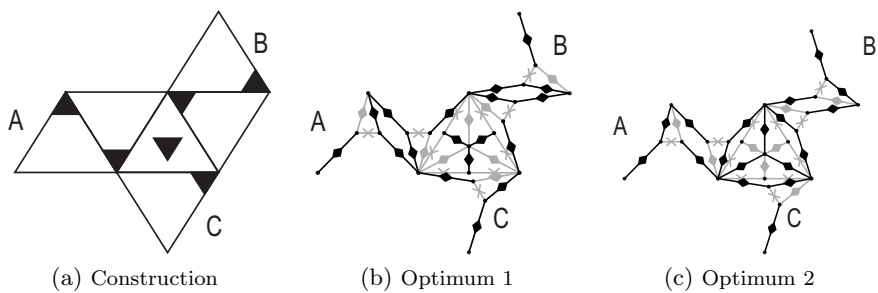

| (a) Construction | (b) Optimum 1 | (c) Optimum 2 |

Figure 8: Edge sum construct.

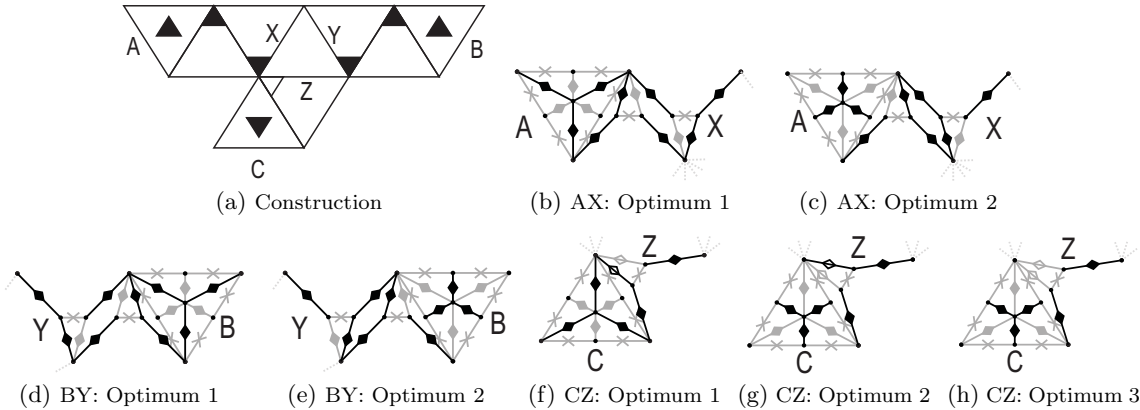

(a) Construction  (b) AX: Optimum 1  (c) AX: Optimum 2

(d) BY: Optimum 1  (e) BY: Optimum 2  (f) CZ: Optimum 1  (g) CZ: Optimum 2  (h) CZ: Optimum 3

Figure 9: Edge sum construct.

By construction, if a locally optimal coalition structure exists, then the original 3-SAT problem must be satisfiable. Furthermore, if the 3-SAT problem is satisfiable, then we can simply set each construct to the locally optimal coalition structure that agrees with the logical value of the variables and their literals, and create a coalition structure for the entire graph by taking the union of any overlapping coalitions. Note, this is always possible by construction. The constructs in Figures 7 to 9 are designed so that under the induced optimums, the nodes in A are never in the same coalition as a node from B or C, and the nodes in B are never in the same coalition as a node from C. Moreover, the construct in Figure 6 is such that in a locally optimal structure, coalition XY is always disjoint from the nodes in A and B. This means that combining two locally optimal coalition structures that agree across such pairs will only create coalitions that are local to the two pairs of nodes being connected and the edges used to connect them. Thus, combining over several such connections is always possible without contradiction.

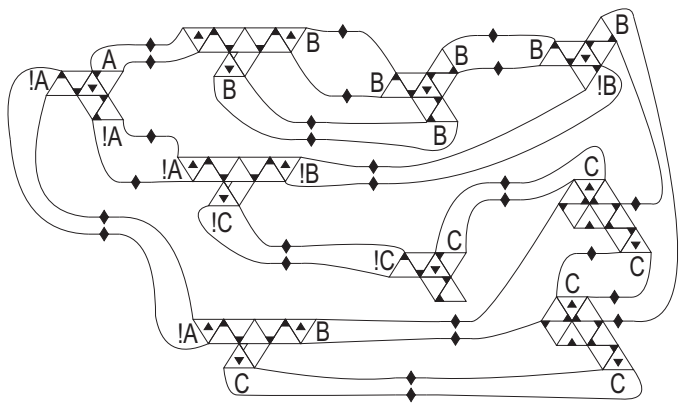

Figure 10: Reduction of $(A \vee B \vee B) \wedge (!A \vee !B \vee !C) \wedge (!A \vee B \vee C)$.

So, a locally optimal coalition structure exists if and only if the original 3-SAT problem is satisfiable, and given any locally optimal coalition structure, we can identify a solution to the 3-SAT problem. Furthermore, if a locally optimal coalition structure exists, then a coalition structure is optimal if and only if it is locally optimal. The size of this graph is $O(m^2)$ and thus the result follows. An example of this reduction process is shown in Figure 10 for the 3-SAT problem $(A \vee B \vee B) \wedge (!A \vee !B \vee !C) \wedge (!A \vee B \vee C)$. □

## 6. Related Work

In this section, we give an overview of the related work, that can be broadly classified under two main categories: clustering algorithms and algorithms for coalition structure generation (CSG). The former is relevant to this work because it deals with partitioning graph structures into subgraphs; however, unlike in our case, the values of such partitions are determined by a certain, problem–specific valuation function. The latter, on the other hand, considers more general valuation functions, but allows no structure on the primitive set of elements.

In more detail, clustering is one of the primery tools in machine learning that deals with finding a structure in a collection of unlabeled data. The goal is to organise objects into groups—*clusters*—whose members are "similar" between them and are "dissimilar" to the objects belonging to other clusters. In certain relevant scenarios, instead of the actual description of the objects, the relationships between them are known. Thus, like in our work, the objects are typically represented by the node set of a signed graph, where the edge labels indicate whether two connected nodes are similar or different. However, clustering algorithms are usually designed for solving problems associated with particular objectives (and hence, valuation functions)—e.g., correlation or modularity that we mentioned in previous sections. In contrast, our work is concerned with a general class of valuation functions, characterised by a single assumption of the independence of disconnected members. Thus, in particular, our Corollary 1 can be viewed as a generalisation of the result by Xin (2011) providing a linear time algorithm for correlation clustering over graphs with bounded treewidth. In this sense, the literature on the CSG problem that we survey below, is perhaps more relevant to our research, as it deals with designing universal algorithms, for which a valuation function is part of an input. However, on the other hand, most of these works assume no structure on the primitive set of elements.

There have been several algorithms developed for CSG. Sandholm, Larson, Andersson, Shehory and Tohmé In (1999), proposed an anytime procedure with worst case guarantees; however, it only reaches an optimal solution after checking all possible coalition structures, and so runs in time $O(n^n)$. Specifically, given a graph where the node set represents coalition structures, which are connected by an edge if and only if they belong to two consecutive levels such that a coalition structure in level $(i-1)$ can be obtained from the one in level $i$ by merging two coalitions into one, the algorithm firstly searches the two bottom levels, and then explores the remaining levels one by one, starting from the top and moving downwards. A similar algorithm was proposed by Dang and Jennings (2004): after searching the two bottom and one top level, the algorithm goes through certain subsets of all remaining levels (as determined by the sizes of coalitions present in their corresponding structures), instead of searching the levels one by one. On the other hand, algorithms based on dynamic programming (DP) (Yeh, 1986; Rothkopf, Pekeč, & Harstad, 1998) work by iterating over all coalition structures of size 1, and then over all those of size 2, and so on until size $n$: for every such coalition $C$, the value of the coalition is compared to the value that could possibly be obtained by splitting $C$ into two coalitions. Visualising such a process with the graph of coalition structures as before, we start from the bottom node and move upwards through a series of connected nodes (a "path") until an optimal node is reached. Importantly, if there are multiple paths that lead to the same optimal node, then DP can reach it through any of these paths. Based on this observation, an improved dynamic programming algorithm (IDP) was developed by Rahwan and Jennings (2008b). The main idea of IDP is to remove edges in the coalitions structure graph so that to disregard as many splittings of coalitions as possible, yet without losing the guarantee of having a path that leads to every node in the graph. This avoids counting approximately 2/3 of the operations compared to DP that evaluates every edge in the coalition structure graph, meaning IDP can find an optimal solution in $O(3^n)$ time. However, DP and IDP algorithms are not anytime—that is, they do not allow to trade computation time for solution quality. To this end, Rahwan, Ramchurn, Giovannucci and Jennings (2009) developed the integer partition (IP) algorithm, which is anytime. It works by dividing the search space into regions, according to the coalition structure configurations based on the sizes of coalitions they contain, and then performing branch-and-bound search. Although this procedure has the worst case complexity of $O(n^n)$, in practice, it is much faster than the DP based algorithms. Furthermore, the IP algorithm was improved upon, by using DP for preprocessing (Rahwan & Jennings, 2008a). To date, this combined algorithm, termed IDP-IP, is the fastest anytime algorithm, that is capable of finding an optimal solution in $O(3^n)$ time.

The CSG problem has also been tackled with heuristic methods. In particular, Sen and Dutta (2000) gave a genetic algorithm that starts with an initial, randomly generated, set of coalition structures, called a "population", and then repeatedly evaluates every member of the current population, selects members based on their evaluation, and constructs new members from the selected ones by exchanging and/or modifying their contents. Keinnen (2009), based the process on Simulated Annealing—a generic, stochastic local search technique: at each iteration, the algorithm explores different neighbourhoods of a certain coalition structure, where every neighbourhood is defined according to a different criterion.

On the other hand, Shehory and Kraus (1998) proposed a decentralised greedy procedure where at each iteration, the best of all candidate coalitions (those that do not overlap with coalitions currently present in the coalition structure) is added to the structure, and the search is done in a distributive fashion—i.e., the agents negotiate over which one of them searches which coalitions. A significantly improved distribution mechanism was later on proposed by Rahwan and Jennings (2007). Another greedy algorithm (Mauro, Basile, Ferilli, & Esposito, 2010) is based on GRASP—a general purpose greedy algorithm that, after each iteration, performs a quick local search to try and improve its solution (Feo & Resende, 1995). In the CSG version of GRASP, a coalition structure is constructed iteratively, where every iteration consists of two steps: the first is to add the best candidate coalition to the structure, and the second is to explore different neighbourhoods of the current structure. These two iterations are repeated until the whole set of agents is covered, and then the whole process is repeated to achieve better solutions. However, all these heuristic techniques do not guarantee that the optimal value will be reached at any point, nor do they give the means of evaluating the quality of the coalition structure selected.

An alternative approach to the CSG problem is to utilise compact representation schemes for valuation functions proposed (Ohta, Conitzer, Ichimura, Sakurai, Iwasaki, & Yokoo, 2009). Indeed, in practice, these functions often display significant structure, and there have been several methods developed to represent them concisely (e.g., by a set of "rules" to compute the function or in terms of "skills" possessed by the agents or their "types" determining their possible contribution to a coalition). Thus, for *marginal contribution nets*, or *MC-nets* (Ieong & Shoham, 2005), the CSG problem was formulated as a mixed integer program (MIP) (Ohta et al., 2009), which can be solved reasonably well compared to the IP algorithm, which does not make use of compact representations. However, in general the problem stays NP-hard, which was also shown for other compact representations such as *synergy coalition groups* (Conitzer & Sandholm, 2006) and *skill games* (Ohta, Iwasaki, Yokoo, Maruono, Conitzer, & Sandholm, 2006; Bachrach, Meir, Jung, & Kohli, 2010) (for the latter, the authors were also able to define a subclass of instances in which the problem can be solved in time polynomial in the number of agents $n$ and the number of skills $k$). For *agent-type* representation, two dynamic programming algorithms were proposed to solve the CSG problem (Aziz & de Keijzer, 2011; Ueda, Kitaki, Iwasaki, & Yokoo, 2011), and both run in $O(n^{2t})$ time, where $t$ is the number of different types.

Another interesting direction was to look at coalition structure generation in the framework of *distributed constraint optimisation problems* (DCOPs) that has recently become a popular approach for modeling cooperative agents (Modi, 2003). Thus, Ueda, Iwasaki, Yokoo, Silaghi and Matsui (2010) consider the CSG problem in a multi-agent system represented as one big DCOP, where every coalition's value is computed as the optimal solution of the DCOP among the agents of that coalition. Instead of solving $O(2^n)$ DCOPs, the authors suggest modifying the big DCOP and solving it using existing algorithms, e.g., ADOPT (Modi, 2003) or DPOP (Petcu & Faltings, 2005).

On the other hand, Rahwan, Michalak, Elkind, Faliszewski, Sroka, Wooldridge and Jennings (2011) proposed the constrained coalition formation (CCF) framework, where there are constraints on the coalition structures that can be formed. In particular, a CCF problem is given by a set of agents, the set of feasible coalition structures and the characteristic function assigning values to coalitions that appear in some feasible coalition structures.

Although in the general case, the notion of feasibility is defined for coalition structures, in many settings of interest the constraints implied on coalition structures can be reduced to constraints on individual coalitions—such settings are termed *locally constrained*. To represent the constraints succinctly, the authors propose the use of propositional logic. They then define a natural subclass of locally constrained CCF problems for which they develop an algorithm to solve the CSG problem which is based on divide-and-conquer techniques.

Finally, a couple of recent papers considered the problem of coalition structure generation on combinatorial structures—i.e., graphs. Thus, Aziz and de Keijzer (2011) showed polynomial time bounds for coalition structure generation in contexts of spanning tree games, edge path coalitional games and vertex path coalitional games, where the value of a coalition of nodes is either 1 or 0, depending on whether or not it contains a spanning tree, an edge path or a vertex path, respectively. The authors also prove NP-hardness of the GCSG problem on general graphs with the edge sum valuation function. In this paper, we present a stronger result showing the hardness of the problem for planar graphs. Independently, Bachrach, Kohli, Kolmogorov and Zadimoghaddam (2011) showed that the coalition structure generation problem is intractable for planar graphs with the edge sum valuation function, and also provided algorithms with constant factor approximations for planar, minor–free and bounded degree graphs. However, in both aforementioned papers, like in the classic literature on clustering, the problem is considered in a particular context (i.e., is associated with a specific valuation function). In contrast, the results presented here apply to a general class of valuation functions, characterised by a single assumption of the independence of disconnected members.

## 7. Conclusions

A key organisational form in multi-agent systems involves members of the same coalition coordinating their actions to achieve common goals. If the agents are organised effectively, their cooperation can significantly improve the performance of each individual and a system as a whole, especially in cases where single agents have insufficient skills or resources to complete the given tasks on their own. For this reason, generating good coalitional structures is one of the fundamental problems studied in AI.

However, in many real-life scenarios, only certain subsets of agents are able to cooperate and apply joint actions. Indeed, to act collectively, a group of agents has to 1) find a (most) beneficial plan of action, 2) agree on it, and 3) coordinate actions among the members of the group. Now, this may not be achievable by an arbitrary subset of agents which are not connected or related to each other. Therefore, the study of coalition formation while taking into consideration the social (or, communication) structure of the set of participants, besides being a most natural and interesting research direction, may provide a key to many positive results in terms of the problem tractability, as well as the quality and stability of solutions. Moreover, this approach is obviously much more appealing from a practical perspective than that of considering agents as interacting in a "vacuum".

To this end, this paper studies the problem of coalition structure generation over graphs (GCSG) and provides the foundation for analysis of its computational complexity. Our work stands out from the existing literature on graph partitioning (or, clustering) in that it does not focus on a specific coalition valuation function, but rather looks at a general class of functions characterised by a single assumption of the independence of disconnected members (IDM).

Our results show that in certain important cases it is indeed valuable to identify that the valuation function satisfies the IDM property, as this significantly reduces the complexity of the GCSG problem one faces. In particular, Algorithm 1 uses a simple search procedure with a guaranteed bound of $O\left(n^2\binom{e+n}{n}\right)$ computational steps for general graphs with $n$ nodes and $e$ edges. Hence, whenever the graph is sparse so that this bound gets lower than $3^n$—the number of steps required to solve the coalition structure generation problem over an unstructured set of elements—utilising the graph structure is beneficial. For a graph with $n$ nodes and a known tree decomposition of width $w$, Algorithm 2 requires $O(w^{w+O(1)}n)$ computational steps, implying that the problem can be solved in linear time for bounded treewidth graphs! In addition, coupling Algorithm 2 with existing separator theorems for minor–free and planar graphs, provides improved computational bounds for coalition structure generation over these important graph classes, although, as we show in Theorem 2, the problem remains NP–complete even for planar graphs with simple edge sum valuation functions.

Our work suggests several directions for future research on this topic. First, although the theoretical bounds we give on complexity of the problem on minor–free and planar graphs are close to best possible, they are not tight. Closing this gap would complete our results. Second, and perhaps the main direction in this study, is exploring the approximability of the GCSG problem for these and other interesting graph classes, and developing approximation schemes where applicable. In this line, partial results are provided by Bachrach, Kohli, Kolmogorov and Zadimoghaddam (2011) that give algorithms with constant factor approximations for planar, minor–free and bounded degree graphs endowed with the edge sum valuation function. It is a challenging task to see if and how these results extend to a more general class of the IDM functions. Finally, it would be interesting to incorporate the ideas of compact representation (Ohta et al., 2009) and constrained coalition formation (Rahwan et al., 2011) into graph coalition structure generation.

## Appendix A

**Lemma 1** *Given a graph $G = (N, E)$ and a coalition valuation function $v(\cdot)$ with the IDM property, for any $A, B \subseteq N$ if there are no edges in $G$ between $A \setminus B$ and $B \setminus A$, then*

$$v(A) - v(A \cap B) = v(A \cup B) - v(B). \tag{3}$$

**Proof :** If $B \setminus A = \emptyset$ then $A \cap B = B$ and $A \cup B = A$, so the result holds. Now, let us show that it holds when $\|B \setminus A\| = 1$. Suppose otherwise, then let $A$ and $B$ be such that $\|A \setminus B\|$ is minimal over $A$ and $B$ where $\|B \setminus A\| = 1$ and (3) is violated. We cannot have $A \setminus B = \emptyset$, for otherwise $A \cap B = A$ and $A \cup B = B$, which would imply that (3) holds. Let $x$ be some element of $A \setminus B$. Then, from the IDM property,

$$v(A) - v(A \setminus \{x\}) = v(A \cup B) - v((A \cup B) \setminus \{x\}), \tag{4}$$

but, by choice of $A$ and $B$, the set $A \setminus \{x\}$ must satisfy (3), and since $x$ is not in $B$, we then have

$$v(A \setminus \{x\}) - v(A \cap B) = v((A \cup B) \setminus \{x\}) - v(B). \qquad (5)$$

Adding up (4) and (5) gives us that $v(A) - v(A \cap B) = v(A \cup B) - v(B)$, a contradiction.

Now we will show that the result holds in general. Suppose otherwise, then let $A$ and $B$ be such that $\|B \setminus A\|$ is minimal over $A$ and $B$ where (3) is violated. Let $x$ be some element of $B \setminus A$ and let $A' = A \cup (B \setminus \{x\})$. Now, $A' \cap B = B \setminus \{x\}$ and $A' \cup B = A \cup B$. Furthermore, $B \setminus A' = \{x\}$, and so applying the results proven so far for the pair $A'$, $B$, we get

$$v(A') - v(A' \cap B) = v(A' \cup B) - v(B),$$

meaning

$$v(A \cup (B \setminus \{x\})) - v(B \setminus \{x\}) = v(A \cup B) - v(B).$$

Furthermore, by choice of $A$ and $B$, and since $x$ is not in $A$,

$$v(A) - v(A \cap B) = v(A \cup (B \setminus \{x\})) - v(B \setminus \{x\}).$$

These two relations prove that the result holds for $A$ and $B$, which is a contradiction. This completes the proof. ☐

**Lemma 2** *Let* $G = (N, E)$ *be a graph with a tree decomposition* $(X, T)$*, where* $X = \{X_1, \ldots, X_m\}$ *for* $m \le n = |N|$ *and* $T$ *is a tree over* $X$*. Suppose further that the* $X_i$ *are numbered in order of shortest distance in* $T$ *from* $X_1$*, where* $X_1$ *may be chosen arbitrarily. Then, for any* $C \subseteq N$,

$$v(C) = \sum_{i=1}^{m} v(C \cap X_i) - v\big(C \cap X_i \cap \bigcup_{j<i} X_j\big). \qquad (6)$$

**Proof :** Towards a contradiction, let us suppose this result does not hold for some $G$. Let $(X, T)$ be the tree decomposition with minimal $m = \|X\|$ such that (6) is violated. If $m = 1$, then $X = \{N\}$ and equation (6) becomes

$$v(C) = v(C \cap N) - v(C \cap N \cap \emptyset),$$

which is trivially true. So we must have $m > 1$. From the choice of numbering, $X_m$ must be a leaf node in $T$. Let $k$ be such that $X_k$ is the only node $X_m$ is connected to in $T$. Since $X_m \setminus X_k$ is disjoint from all $X_i$ with $i \ne m$, there can be no edges in $G$ between elements of $X_m \setminus X_k$ and $X_k \setminus X_m$. Furthermore, for any $i < m$ such that $i \ne k$, $X_m \cap X_i \subseteq X_m \cap X_k$, and so $X_m \cap \bigcup_{j<m} X_j = X_m \cap X_k$ and

$$X_k \cap \bigcup_{j<k} X_j = (X_m \cup X_k) \cap \bigcup_{j<k} X_j.$$

Thus, for any $C \subseteq N$,

$$v(C \cap X_m) - v(C \cap X_m \cap \bigcup_{j<m} X_j) + v(C \cap X_k) - v(C \cap X_k \cap \bigcup_{j<k} X_j)$$

$$= v(C \cap X_m) - v(C \cap X_m \cap X_k) + v(C \cap X_k) - v(C \cap (X_m \cup X_k) \cap \bigcup_{j<k} X_j)$$

$$= v(C \cap (X_m \cup X_k)) - v(C \cap (X_m \cup X_k) \cap \bigcup_{j<k} X_j),$$

by Lemma 1. Furthermore, for all $i < k$,

$$X_i \cap \bigcup_{j<i} X_j = X_i \cap \big(X_m \cup \bigcup_{j<i} X_j\big),$$

and so,

$$\sum_{i=1}^{m} v(C \cap X_i) - v\big(C \cap X_i \cap \bigcup_{j<i} X_j\big)$$

$$= \sum_{i=1}^{m-1} v(C \cap Y_i) - v\big(C \cap Y_i \cap \bigcup_{j<i} Y_j\big),$$

where $Y_i = X_i$ for $i \neq k$ and $Y_k = X_k \cup X_m$. However, these $Y_i$ form a tree decomposition of $G$ that has only $m - 1$ nodes, (with the tree topology of $T$ with the $X_m$ leaf removed), and thus this sum must equal $v(C)$ by the choice of $m$. Since $C$ was chosen arbitrarily, this leads us to a contradiction, and the result must hold in general. $\qquad\square$

**Lemma 3** For any graph $G = (N, E)$, for any $P, Q \subseteq N$, if $\mathcal{P}$ is a coalition structure over $P$ and $\mathcal{Q}$ is a coalition structure over $Q$, and if $\mathcal{P}(P \cap Q) = \mathcal{Q}(P \cap Q)$, then $\mathcal{E} = U(\mathcal{P}, \mathcal{Q})$ is a coalition structure over $P \cup Q$ and for any $P' \subseteq P$, and $Q' \subseteq Q$, $\mathcal{E}(P) = \mathcal{P}(P')$ and $\mathcal{E}(Q) = \mathcal{Q}(Q')$.

**Proof :** Firstly, for all $A \in \mathcal{P}$, either $A \subseteq (P \setminus Q)$ or there is some $B \in \mathcal{Q}$ such that $A \cap B \neq \emptyset$. Thus, the union of all sets in $\mathcal{E}$ covers all of $P$. By symmetry, the union of all sets in $\mathcal{E}$ must then also cover all of $Q$.

Now, for any $P' \subseteq P$,

$$\mathcal{E}(P') = \{(A \cap P') : A \in \mathcal{P}, A \subseteq (P \setminus Q)\} \cup \{(A \cup B) \cap P' : A \in \mathcal{P}, B \in \mathcal{Q}, A \cap B \neq \emptyset\}.$$

However, for any $A \in \mathcal{P}$, $B \in \mathcal{Q}$ with $A \cap B \neq \emptyset$, since $\mathcal{P}(P \cap Q) = \mathcal{Q}(P \cap Q)$ is a coalition structure over $P \cap Q$, we must have that $A \cap (P \cap Q) = B \cap (P \cap Q)$. As $B \subseteq Q$, $B \cap P'$ is equal to $(B \cap (P \cap Q)) \cap P' \subseteq A \cap P'$. Thus,

$$\mathcal{E}(P') = \{(A \cap P') : A \in \mathcal{P}\} = \mathcal{P}(P').$$

By symmetry, for any $Q' \subseteq Q$, $\mathcal{E}(Q') = \mathcal{Q}(Q')$.

It remains to show that $\mathcal{E}$ is a coalition structure. Towards a contradiction, suppose we have some $A, B \in \mathcal{E}$ such that $A \cap B \neq \emptyset$ and $A \neq B$. Then, since $\mathcal{E}(P) = \mathcal{P}$, and $\mathcal{P}$ is a coalition structure, we must have either $A \cap P = B \cap P$ or $A \cap P$ and $B \cap P$ are disjoint. Likewise, either $A \cap Q = B \cap Q$ or $A \cap B \cap Q = \emptyset$. Now, if $A \cap P = B \cap P$ and $A \cap Q = B \cap Q$, then $A = B$ and if $A \cap B \cap P = \emptyset$ and $A \cap B \cap Q = \emptyset$, then $A \cap B = \emptyset$, both contradictions. Suppose $A \cap P = B \cap P$ and $A \cap B \cap Q = \emptyset$. This implies that $A \cap P = B \cap P$ is non-empty, as $A \cap B \neq \emptyset$, but it also implies that $A \cap P \cap Q = B \cap P \cap Q = \emptyset$, which means, $A \cap P$ is an element of $\mathcal{P}$ that is a subset of $P \setminus Q$. However, the only element of $\mathcal{E}$ that would have $A \cap P$ as a subset would be $A \cap P$ itself, meaning $A = B = A \cap P$, another contradiction. By symmetry, having $A \cap Q = B \cap Q$ and $A \cap B \cap P = \emptyset$ also leads to a contradiction, and therefore this scenario is impossible. Thus, we have shown that $\mathcal{E}$ is a coalition structure, as required. □